\documentstyle[preprint,aps]{revtex}
\begin{document}
\draft
\title{Analytic Continuation of Thermal $N$-Point Functions \\
from Imaginary- to Real-Energies}

\author{R. Baier}
\address{
Fakult\"{a}t f\"{u}r Physik, Universit\"{a}t Bielefeld,
33501 Bielefeld, Germany
}
\author{A. Ni\'{e}gawa}
\address{
Department of Physics, Osaka City University, Sumiyoshi-ku,
Osaka 558, Japan
}
\maketitle
\begin{abstract}
We consider thermal $n$-point Green functions in the framework of
quantum field theory at finite temperature. We show how analytic
continuations from imaginary to real energies relate these functions
originally defined in the imaginary-time formalism to retarded and
advanced real-time ones. The described method is valid to all orders
of perturbation theory. It has the further advantage that it is
independent of approximations often applied in actual finite-order
calculations.
\end{abstract}

\narrowtext
\section{INTRODUCTION}
In quantum field theory at finite temperature and density
two convenient formalisms that enable the use of conventional
Feynman rules in momentum space are applied; the
imaginary-time formalism (ITF) \cite{mat} and the real-time
formalism (RTF) \cite{tak,sch}. (For books and reviews on thermal
field theory, see, e.g. \cite{abr,mil,cho,lan}). The former one is
particularly suited for the evaluation of the static or
thermodynamic quantities of finite-temperature systems, while the
latter is preferred for the evaluation of time-dependent quantities.

Studies of the two-point functions have a long history
\cite{mat,tak,sch,abr,mil,cho,lan,bay}. The relationship between
their representations --- the one obtained from ITF and the other
from RTF --- is well known \cite{bay}. On the other hand the
interest in the three- and $n(\geq 4)$-point functions started
rather recently \cite{bra,bai,eva1,kob,eva2,aur}.

An apparent difference \cite{bai} between the results for
three-point functions obtained from ITF and from RTF has posed
the question: through analytic continuations of the Green functions
evaluated in ITF, what kind of (combinations of) thermal Green
functions in RTF are obtained?

Several papers have been devoted to this issue. It has been
shown, either to one-loop order or in all orders of
perturbation theory, that the three-point function in
ITF, when analytically continued to real energies, becomes the
retarded or advanced three-point Green function \cite {eva1,kob}.
Recently, the $n \; (\geq 4)$-point functions have been investigated
\cite{eva2,aur} and it is shown that different analytic
continuations of the ITF result yield different
RTF Green functions, including the retarded and advanced Green
functions.

The purpose of this paper is to show that the most
straight\-for\-ward an\-a\-lyt\-ic con\-tin\-u\-a\-tion of the
non-am\-pu\-tat\-ed $n$-point Green func\-tion in ITF leads to the
re\-tard\-ed or ad\-vanc\-ed $n$-point Green func\-tion.

In Sec. II, we introduce the $n$-point thermal Green functions with
time arguments on a --- to a large extent arbitrary --- contour
in the complex time plane. We then formulate the problem
of analytic continuation of the Green function in ITF from
imaginary- to real-energies. For the purpose of illustrating our
procedure, we discuss in Sec. III the analytic continuation
of the two-point function. In Sec. IV, we carry out the analytic
continuation of the $n$-point Green functions evaluated in ITF, and
obtain the retarded- and advanced-Green functions. Sec. V is devoted
to conclusions.
\section{Preliminaries}

Throughout this paper we consider a real scalar
field $\phi(x)$. Generalizations to other kind of fields are
straightforward. The thermal Green functions are defined as the
statistical average of a product of Heisenberg fields,
\begin{eqnarray}
G( \{ t \} ) & = &   G(t_1, t_2, \cdots , t_n)  \nonumber \\
& \equiv & Tr \left[ e^{- \beta H} T_c
\left( \phi (t_1) \phi (t_2)  \cdots \phi (t_n) \right) \right]
/ Tr e^{- \beta H} \nonumber \\
  & \equiv & \langle T_c \left( \phi (t_1) \phi (t_2)  \cdots
  \phi (t_n) \right) \rangle \;,
\end{eqnarray}
where $\beta =T^{-1}$ is the inverse temperature and $H$ is the
Hamiltonian of the system such that $\phi(t)=e^{iHt} \phi(0)
e^{-iHt}$. In Eq. (1) and in the following we suppress explicit
reference to the space variables. The arguments $t_1, t_2, \cdots ,
t_n$ lie on the contour $C$ running from $t_0$ to $t_0 - i \beta$
in the complex time plane. The symbol $T_c$ in (1) is the
time-ordering operator along this path $C$. That is, it prescribes
that the operators it is applied to be arranged in the order in
which their arguments lie along $C$, with those nearest to the
beginning at $t_0$ to the right, and those nearest to the end
$t_0 - i \beta$ to the left.

To perform the thermal trace (1), we insert a complete set of states
by choosing the Hamiltonian eigenstate basis:
\widetext
\begin{eqnarray}
G(\{ t \}) Tr e^{- \beta H} & = & \sum_{p_n}
\left[ \prod_{j=1}^{n-1}
\theta_c (t_{\overline j} - t_{\overline {j+1}} ) \right]
\sum_{l_1,l_2, \cdots, l_n}
exp \! \left( -iE_{l_1} (t_{\overline n} - t_{\overline 1}
- i \beta)  \right) \nonumber \\
 & & \times \langle l_1 | \phi (0) | l_2 \rangle \prod_{j=2}^n
 \left[ exp \! \left( i E_{l_j} (t_{\overline j}
 - t_{\overline {j-1}})  \right) \langle l_j | \phi (0) |
 l_{j+1} \rangle \right]  \; ,\; \; \; \; \; \;
 (l_{n+1} \equiv l_1) \;,
\end{eqnarray}
\narrowtext
\noindent with $\theta_c$ the contour step function. The summation
here is carried out over all permutations of $n$ numbers $p_n$
\begin{equation}
   \left(
   \begin{array}{cccc}
   1 & 2 & \cdots & n \\
   {\overline 1} & {\overline 2} & \cdots & {\overline n}
   \end{array}
   \right) \; .
\end{equation}

Following common practice, we assume that the convergence of the
above trace sum (2) is controlled by the exponential factors, so
that it converges if, for every pair of $t_i$ and $t_j$  such that
$\theta_c (t_i - t_j)=1$, $\Im (t_i - t_j)<0$, and
$\Im (t_s - t_l)<\beta$ with $t_s \; (t_l)$ the ``smallest''
(``largest'') time. This condition guarantees the existence of
$G(\{t\})$ (Eq. (2)) as an analytic function of
$\{t\}=\{t_1,t_2, \cdots , t_n\}$. The limit of an analytic function
on the boundaries of its domain of definition, where it is still
continuous, is a generalized function. This implies that the thermal
Green function $G(\{t\})$ is well defined for $\Im (t_i -t_j)
\leq 0$ when $\theta_c (t_i -t_j)=1$, and $\Im (t_s -t_l)
\leq \beta$. This imposes the restriction on $C$ that as a point
moves along $C$ from $t_0$ to $t_0 - i \beta$ its imaginary part is
nonincreasing. Then, an analytic continuation of $G(\{t\})$ can be
done by deforming the contour $C$ with the end points $t_0$ and
$t_0 - i \beta$ held fixed, keeping in mind the
\lq\lq nonincreasing'' condition for the imaginary part of the
points on $C$.

Among the above class of paths, we consider first a special path
$C_I$ depicted in Fig. 1, which defines the ITF. We note that we can
choose any value for $\Re \, t_0$ because of the property of
time-translation invariance of (1). Next we evaluate the Fourier
component of (1),
\begin{eqnarray}
\lefteqn{
G(\{\omega \}) = G(\omega_1,\omega_2, \cdots , \omega_n)
}
\nonumber \\
& & \mbox{\hspace{11.2mm}}= \prod_{j=1}^n \left( \int_{C_I} d t_j \,
e^{- \omega_j t_j} \right)
G(t_1, t_2, \cdots, t_n), \nonumber \\
& & \mbox{\hspace{11ex}} \omega_j = 2 \pi l_j/\beta \, ,
\mbox{\hspace{2ex}} l_j =0, \pm 1, \pm 2,  \cdots, \pm \infty \;.
\end{eqnarray}
It is to be noted that real (discrete) $\omega_j$'s here are what
we call imaginary energies. By using once more the time translation
invariance we rewrite (4) as
\begin{mathletters}
\begin{equation}
\mbox{\hspace{-9.4mm}} G(\{\omega\}) = - i \beta \,
\delta ( \sum_{j=1}^n \omega_j \,
; \, 0) \,
\tilde{G} (\omega_2, \omega_3, \cdots, \omega_n) \; ,
\end{equation}
\begin{eqnarray}
\lefteqn{
\tilde{G} (\omega_2, \omega_3, \cdots, \omega_n)
} \nonumber \\
 & & \mbox{\hspace{3ex}} = \prod_{j=2}^n
 \left( \int_{t_0 -t_1}^{t_0 -t_1 - i \beta}
dt_j \, e^{ - \omega_j t_j }   \right) G(0, t_2, \cdots, t_n)\;.
\end{eqnarray}
\end{mathletters}
\narrowtext
\noindent Here the integrations are performed along the
im\-ag\-i\-na\-ry-time axis. In (5a)
$\delta ( \cdots \, ; \, \cdots )$ denotes the Kronecker $\delta$
symbol. In order to obtain (5) the KMS condition \cite{kub}, which
represents the invariance of the trace under the following cyclic
permutation (cf. (1)),
\begin{equation}
\langle \phi (t_1) \cdots  \phi (t_{n-1}) \phi (t_n) \rangle =
\langle \phi (t_n - i \beta) \phi (t_1) \cdots \phi (t_{n-1})
\rangle \;
\end{equation}
is used. In ITF one calculates $G$ or $\tilde{G}$ as defined
in (5). In the following we focus our attention on how to continue
(5) from imaginary to real energies.
\section{Two-point thermal Green function}

Although the relation between ITF and RTF is well understood
\cite{bay} for two-point thermal Green functions, we start with
the thermal two-point Green function in ITF for the purpose of
illustrating our procedure:
\begin{eqnarray}
\lefteqn{
G(\omega_1, \omega_2)
} \nonumber \\
 & & \mbox{\hspace{3ex}} = \int_{C_I} dt_1 \int_{C_I} dt_2 \,
e^{  - (\omega_1 t_1 + \omega_2 t_2 )}
\langle T_{C_I} (\phi (t_1) \phi (t_2) ) \rangle \; .
\end{eqnarray}
As explained in Sec. II, the integrand of (7) is an analytic
function of $t_1$ and $t_2$ in the strip,
$\Im \, t_0 - \beta< \Im \, t_j < \Im \, t_0 \;(j=1,2)$ with
$t_1  \neq t_2$, and we may deform the contour $C_I$ keeping the
property as mentioned above after (3). In this way we may obtain
the contour $C_R = C_1 \oplus C_2 \oplus C_3 \oplus C_4$ as depicted
in Fig. 2: the path goes down from $t_0$ to $t_I \equiv \Re \, t_0$,
continues from $t_I$ to $t_F$ along the real axis, returns back to
$t_I$ along the real axis, and ends at $t_0 - i \beta$.

Because of time translation invariance $G(\omega_1, \omega_2)$
is nonvanishing only for $\omega_1 + \omega_2 = 0$ (cf.(5a)), and
the integrand of (7) --- on the path $C_R$ --- is a function of
$t_2 - t_1$. This, together with the above mentioned analyticity
property of $G(t_1,t_2)$, allows us to evaluate
$G(\omega_1, \omega_2)$ by fixing $t_1$ on any point on the contour
$C_R = C_1 \oplus C_2 \oplus C_3 \oplus C_4$. We first fix $t_1$ on
$C_1$, to obtain
\widetext
\begin{mathletters}
\begin{eqnarray}
G(\omega_1, \omega_2) & = & -i \beta \,
\delta (\omega_1; - \omega_2) \left[ \int_{C_3}dt_2 \,
e^{- \omega_2 (t_2 - t_1)} \langle \phi (t_1) \phi (t_2) \rangle
\right. \nonumber \\
& & \mbox{\hspace{2ex}} \left. + \int_{C_1} dt_2 \,
e^{- \omega_2 (t_2 - t_1)}
\langle   T_{C_1} (\phi(t_1) \phi(t_2) ) \rangle +
\int_{C_2 \oplus C_4} dt_2 \, e^{- \omega_2 (t_2 - t_1)}
\langle  \phi(t_2) \phi(t_1)  \rangle  \right]
\end{eqnarray}
\begin{eqnarray}
& &  \mbox{\hspace{-12.4mm}} = -i\beta \,
\delta (\omega_1 ; - \omega_2)
\int_{t_I}^{t_F} dt_2 \, e^{- \omega_2 (t_2 - t_1)}
\left[ \langle  T(\phi (t_1) \phi (t_2)) \rangle  -
\langle  \phi (t_2) \phi (t_1) \rangle  \right] \nonumber \\
& & \mbox{\hspace{-4mm}} + \mbox{(contributions from
$C_3$ and $C_4$)}\;.
\end{eqnarray}
\end{mathletters}
\narrowtext
\noindent The symbol $T$ is the ordinary time-ordering symbol. Eq.
(8) is now well suited for the purpose of an analytic continuation
of $G(\omega_1, \omega_2)$ to real energies. In order to arrive at
RTF with real energy $p_{20} (= - p_{10})$ we take the limit
$t_I (=\Re \, t_0) \to - \infty$ and $t_F \to + \infty$, taking into
account that $t_I (= \Re \, t_0)$ may be chosen arbitrarily.

We realize (cf. \cite{cho}) that the term in the square bracket in
(8b) may be written as
\begin{eqnarray}
\lefteqn{
\langle  T(\phi(t_1) \phi(t_2)) \rangle  -
\langle  \phi(t_2) \phi(t_1) \rangle
} \nonumber \\
& & \mbox{\hspace{15ex}} = \theta (t_1 -t_2) \langle  [ \phi(t_1),
\phi(t_2) ] \rangle \;.
\end{eqnarray}
Then, from (8) and (9), we learn that the above limit,
$t_I \to - \infty$ and $t_F \to + \infty$, can be taken if we
continue to the real energy as follows,
\begin{mathletters}
\begin{equation}
\omega_2 \to - i (p_{20} - i \epsilon )\;,
\end{equation}
or equivalently,
\begin{equation}
\omega_1 \to - i (p_{10} + i \epsilon )\;,
\end{equation}
\end{mathletters}
\narrowtext
\noindent with $\epsilon$ an infinitesimal positive number.

The time path for $G$ consists of two real-time segments, namely
$C_1$ from $- \infty \to + \infty$ and $C_2$ from $+ \infty$ to
$- \infty$, and of $C_3 \oplus C_4$ ($C_3$ from $- \infty + i
\Im \,t_0$ to $- \infty$ and $C_4$ from $- \infty$ to
$- \infty +i( \Im \, t_0 - \beta))$. Then, in evaluating (8) with
(10) by perturbative methods, there are four kind of bare thermal
propagators, $G_{ij}^{(0)}(t_i, t_j) \; (i,j=1-4)$, present with
$t_i \in C_i$.  Following \cite{nie}, one can show that as far as
the thermal Green functions (like (8) with (10)) with their time
arguments lying on $C_1$ and/or $C_2$ are concerned, the net effects
of the contributions from $C_3$ and $C_4$ are to modify
$G_{ij}^{(0)} \; (i,j=1,2)$ to the familiar forms \cite{foo} in
two-component real-time thermal field theory constructed on the
basis of the closed-time path $C_1 \oplus C_2$ \cite{sch,mil,cho}.

The analytic continuation of the energy conservation
$\delta$-function in (8) is given by the prescription \cite{lan}:
$\delta (\omega_1; -\omega_2) \rightarrow 2 \pi i \beta^{-1} \delta
(p_{10}+p_{20})$.

Thus we obtain the analytically continued $G$ with real energies
$p_{10}$ and $p_{20}$,
\begin{mathletters}
\begin{equation}
\mbox{\hspace{-6ex}}
\lim_{\epsilon \rightarrow + 0}
 G (p_{10} + i \epsilon, \, p_{20}
- i \epsilon)  =  G_{++} - G_{+-}
\end{equation}
\begin{eqnarray}
\mbox{\hspace{4ex}}& &  =  2 \pi  \delta (p_{10} + p_{20})
\nonumber \\
& &  \mbox{\hspace{1ex}} \times \lim_{\epsilon \rightarrow + 0}
 \int_{- \infty}^\infty dt \, e^{i (p_{20} - i \epsilon) t}
\, \theta (-t) \langle  [ \phi(0), \phi(t)] \rangle \; ,
\end{eqnarray}
\end{mathletters}
\narrowtext
\noindent where we follow e.g. \cite{cho} and introduce
\begin{eqnarray}
& & G_{\alpha \beta}(p_{10}, p_{20}) \nonumber \\
& & \mbox{\hspace{2ex}} = \int_{- \infty}^{\infty} dt_1
\int_{- \infty}^{\infty} dt_2 \, e^{ i(p_{10}t_1 + p_{20}t_2 )}
\langle  A_{\alpha \beta} (t_1, t_2)
\rangle \; ,
\end{eqnarray}
with the definitions
\begin{eqnarray}
A_{++} (t_1, t_2) & = & \theta (t_1 -t_2) \phi (t_1) \phi(t_2)
+ \theta (t_2 -t_1) \phi (t_2) \phi(t_1) \nonumber \\
A_{+-} (t_1, t_2) & = &  \phi (t_2) \phi(t_1) \nonumber \\
A_{--} (t_1, t_2) & = & \theta (t_1 -t_2) \phi (t_2) \phi(t_1)
+ \theta (t_2 -t_1) \phi (t_1) \phi(t_2) \nonumber \\
A_{-+} (t_1, t_2) & = &  \phi (t_1) \phi(t_2) \;.
\end{eqnarray}
The functions $G_{\alpha \beta}\; (\alpha, \beta =+,-)$ in (12)
denote Green functions in the two-component real-time thermal field
theory, applying the closed time-path formalism introduced by
Schwinger and Keldysh \cite{sch}. It is to be noted, in passing,
that the $A_{\alpha \beta}$'s are not independent since they satisfy
the relation \cite{cho},
\begin{eqnarray}
& & A_{++}(t_1,t_2)+A_{--}(t_1,t_2) \nonumber \\
& & \mbox{\hspace{11ex}} -A_{+-}(t_1,t_2)-A_{-+}(t_1,t_2)=0
\;.
\end{eqnarray}
It is important to note that the continued function $G$ in (11b)
is identical with the retarded Green function. In this way the
\lq\lq physical'' representation as discussed in \cite{cho} is
established, where this function is denoted by
$G=i G_{21}(p_{10}, p_{20})$, however, the primary quantities that
are calculated directly in real-time thermal field theory are
$G_{\alpha \beta}$ in (12).

When we fix $t_2$, instead of $t_1$, on $C_1$, and repeating similar
steps as above we obtain the retarded Green function {\em with
respect to $\phi(t_2)$}, i.e. the advanced Green function with
respect to $\phi(t_1)$. Fixing the time variable as $t_1 \in C_2$
leads to the same result as above, (11), and likewise for the choice
$t_2 \in C_2$. Fixing $t_1$ on $C_3$ or $C_4$, $t_1 \in C_3
\oplus C_4$, is not suitable for analytic continuations under
consideration and we recover the original formula (7); this holds
likewise for the choice $t_2 \in C_3 \oplus C_4$.

Finally we may deform the contour $C_I$ in (7) to the one that is
mirror symmetric to the one of Fig. 2. Changing the time variable
$t_j$ as $t_j \equiv  \Re \, t_j + i \Im \, t_j \to - \Re \, t_j
+ i \Im \, t_j$, we get back the time path $C_1 \oplus C_2 \oplus
C_3 \oplus C_4$ of Fig. 2.  Then, fixing $t_1$ on the upper path on
the real axis in the complex time-plane, we deduce
\begin{mathletters}
\begin{equation}
\mbox{\hspace{-6ex}}
\lim_{\epsilon \rightarrow + 0}
 G (p_{10} - i \epsilon, \, p_{20}
+ i \epsilon)  =  -G_{++} + G_{-+}
\end{equation}
\begin{eqnarray}
\mbox{\hspace{5ex}}& &  =  - 2 \pi  \delta (p_{10} + p_{20})
\nonumber \\
& &  \mbox{\hspace{2ex}} \times \lim_{\epsilon \rightarrow + 0}
 \int_{- \infty}^\infty dt \, e^{i (p_{20} + i \epsilon) t}
\, \theta (t) \langle  [ \phi(0), \phi(t)] \rangle \; ,
\end{eqnarray}
\end{mathletters}
\narrowtext
\noindent i.e. the advanced Green function.
\section{Thermal $n$-point Green function}

We proceed in a similar manner as in Sec. III. Starting with the
thermal $n$-point function in ITF, Eq. (4), we deform the contour
$C_I$ to $C_R$ as depicted in Fig. 2. Under the constraint $\sum_j
\omega_j=0$ (cf. (5a)), we evaluate (4) with $C_R$ for $C_I$ by
fixing $t_1$ on $C_1$, $t_1 \in C_1$, which is suitable for
continuation to real energies. In place of Eq. (8) we now have
\widetext
\begin{mathletters}
\[
\mbox{\hspace{-23ex}}
G(\{ \omega \})  =  - i \beta \, \delta ( \sum_j \omega_j ; 0)
\prod_{j=2}^n \left( \int_{C_1 \oplus C_2} dt_j \,
e^{- \omega_j(t_j-t_1)} \right)
\]
\begin{equation}
\mbox{\hspace{8ex}} \times \langle T_{C_1 \oplus C_2} (\phi (t_1)
\cdots \phi (t_{n-1})  \phi(t_n)) \rangle + \cdots \cdots
\mbox{\hspace{5ex}}  (t_1 \in C_1)\;,
\end{equation}
\begin{eqnarray}
& &  \mbox{\hspace{24.1mm}}
= - i \beta \, \delta (\sum_j \omega_j; 0)
\prod_{j=2}^n \left( \int_{t_I}^{t_F} dt_j \;
e^{- \omega_j (t_j - t_1) } \right)  \nonumber \\
& & \mbox{\hspace{30.7mm}}\times \left[ \sum_{\alpha_2,\alpha_3,
\cdots, \alpha_n =+,-} (-)^s \langle
T_{C_1 \oplus C_2 } \left( \phi_+(t_1) \phi_{\alpha_2}(t_2) \cdots
\phi_{\alpha_n}(t_n) \right) \rangle  \right] + \cdots \cdots \, ,
\end{eqnarray}
\end{mathletters}
\narrowtext
\noindent where
\begin{equation}
s= \frac{1}{2} \sum_{j=2}^n (1 - \alpha_j)\;,
\end{equation}
and the notation of \cite{cho} is used: for $n>2$ it becomes more
transparent to make the subscripts $+,-$ explicit, which indicate
that the times $t_i$ assume values on either the \lq\lq positive''
time path $C_1$ or on the \lq\lq negative'' one $C_2$; e.g. this
implies for all $+$ subscripts time ordering, whereas for all $-$
subscripts anti-time ordering. This generalizes the expressions in
(13) to the $n \geq 3$ cases (for n=3 explicit expressions are given
in \cite{cho}). In Eq. (16) the dots indicate the contributions when
some of the $t$'s among $t_2, \cdots, t_n$ are on $C_3 \oplus C_4$;
this portion is treated according to the arguments given in Sec.III
\cite{nie}.

Next we transform to the \lq\lq physical'' representation. By
applying the procedure described in \cite{cho}, we express the term
in the square bracket in (16b) as,
\begin{equation}
\sum_{p_{n-1}} \theta (1, \overline{2}, \cdots, \overline{n})
\langle [ \, \cdots [ \phi (t_1), \, \phi (t_{\overline{2}}) ],
\, \phi (t_{\overline{3}}) ], \cdots \, ], \,
\phi (t_{\overline{n}}) ] \rangle \;,
\end{equation}
where $\theta$ is the mul\-ti-step func\-tion de\-fin\-ed by
\linebreak[0] $\theta(1,2, \cdots, n)$ $=$ $\theta(1,2)$
$\theta(2,3)$ $\cdots$ $\theta(n-1,n)$ with $\theta(1,2)$ $=$
$\theta(t_1-t_2)$. Eq. (18) is the gen\-er\-al\-i\-za\-tion of the
two-point func\-tion case Eq. (9). The sum\-ma\-tion here is
car\-ried out over all per\-mu\-ta\-tions of $n-1$ num\-bers
$p_{n-1}$
\begin{equation}
   \left(
  \begin{array}{cccc}
  2 & 3 & \cdots & n \\
  {\overline 2} & {\overline 3} & \cdots & {\overline n}
 \end{array}
 \right) \; .
\end{equation}
Eq. (18) tells us that $t_1$ is the largest time, therefore we are
allowed to take the limit $t_I (= \Re \, t_0) \to - \infty$
and $t_F \to  + \infty$ if we continue $\omega_j$ in (16b) to real
energies as
\begin{mathletters}
\begin{equation}
\omega_j \to - i (p_{j0} - i \epsilon)\;, \; \; \; \; \;
j=2,3, \cdots, n,
\end{equation}
and
\begin{equation}
\omega_1 \to - i (p_{10} + i (n-1) \epsilon)\;.
\end{equation}
\end{mathletters}
\narrowtext
\noindent Then we arrive at the Green function in RTF,
\widetext
\begin{mathletters}
\begin{equation}
\lim_{\epsilon \rightarrow +0}
G( p_{10}+i(n-1) \epsilon,
\, \{ p_{j0}-i \epsilon; \, j=2,3, \cdots, n  \})
=  \sum_{\alpha_2, \cdots, \alpha_n =+,-} (-)^s
G_{+,\alpha_2, \cdots, \alpha_n}
\end{equation}
\begin{eqnarray}
& & \mbox{\hspace{10ex}} =  2 \pi  \,
\delta (\sum_{j=1}^n p_{j0} ) \lim_{\epsilon \rightarrow + 0}
\prod_{j=2}^n \left(  \int_{- \infty}^{\infty}  dt_j
e^{ i (p_{j0} - i \epsilon )(t_j - t_1)} \right)
\nonumber \\
& & \mbox{\hspace{12ex}} \times \sum_{p_{n-1}}
\theta (1, \overline{2}, \cdots, \overline{n})
\langle [\, \cdots [ \phi(t_1), \, \phi(t_{\overline{2}})], \,
\phi(t_{\overline{3}})], \cdots \, ], \, \phi(t_{\overline{n}})]
\rangle  \\
& & \mbox{\hspace{10ex}} \equiv i^{n-1} \tilde{G}_{211 \cdots 1}
(p_{10}, \cdots, p_{n0}) \; .
\end{eqnarray}
\end{mathletters}
\narrowtext
\noindent where $s$ is given in (17). Thus we have derived the
Green function (21c) in the \lq\lq physical'' representation,
denoted by $\tilde{G}_{211 \cdots 1}$ \cite{cho}, which is the
thermal $n$-point retarded function as seen in (21b).  In (21a),
$G_{+,\alpha_2, \cdots, \alpha_n}$ is a Green function in RTF and it
is defined \cite{cho} analogously to (12) and (13),
\begin{eqnarray}
\lefteqn{
G_{\alpha_1,\alpha_2, \cdots, \alpha_n}
} \nonumber \\
& &  \mbox{\hspace{5ex}} = \prod_{j=1}^n \left(
\int_{- \infty}^{\infty} dt_j  e^{i p_{j0} t_j} \right) \nonumber \\
& & \mbox{\hspace{8ex}} \times \langle  T_{C_1 \oplus C_2}
\left(  \phi_{\alpha_1}(t_1)
\phi_{\alpha_2}(t_2)\cdots \phi_{\alpha_n}(t_n)   \right) \rangle
\; .
\end{eqnarray}
As in the two-point function case, (14), there is one identity;
\begin{equation}
\sum_{\alpha_1, \cdots, \alpha_n =+,-} (-)^{s'}
G_{\alpha_1, \alpha_2, \cdots, \alpha_n} = 0
\end{equation}
where
\begin{equation}
    s' = \frac{1}{2} \sum_{j=1}^n (1 - \alpha_j) \; .
\end{equation}
It is worth mentioning that the primary quantities that are
evaluated in real-time thermal field theory are
$G_{\alpha_1,\alpha_2, \cdots, \alpha_n}$ defined in (22), through
which the retarded Green function $G$ in (21b) is
obtained.

We have developed the derivation by fixing $t_1$ on $C_1$.
Of course, we may proceed in a similar manner by fixing other
$t_j \, (2 \leq j \leq n)$ on $C_1$: $n-1$ different retarded Green
functions are the result.

In case Eq.(4) is evaluated with $C_R$ for $C_I$ by fixing $t_1$ on
$C_2$, we obtain the same result as above, (21); for $t_1 \in C_3
\oplus C_4$ we recover the original formula (4).

When we deform the contour $C_I$ in (4) to the one that is mirror
symmetric to the one of Fig. 2, and fixing $t_1$ on the upper path
on the real axis in the complex-time plane, i.e. the counterpart of
$C_1$ in Fig. 2, we derive the advanced Green function,
\widetext
\begin{eqnarray}
& & \mbox{\hspace{-14ex}} \lim_{\epsilon \rightarrow + 0}
G( p_{10} - i (n-1)
\epsilon, \, \{ p_{j0} + i \epsilon; \; j=2,3,
\cdots, n \})
  \nonumber \\
& &  = - 2 \pi \, \delta (\sum_{j=1}^n p_{j0})
\lim_{\epsilon \rightarrow +0}  \prod_{j=2}^n \left(
\int_{- \infty }^{\infty}  dt_j \,
e^{ i(p_{j0} + i \epsilon ) (t_j -t_1)}  \right)  \nonumber \\
& &  \mbox{\hspace{2ex}} \times \sum_{p_{n-1}} \theta(\overline{n},
\cdots, \overline{2}, 1) \langle [\, \cdots [ \phi(t_1), \,
\phi(t_{\overline{2}})], \, \phi(t_{\overline{3}})], \cdots \, ], \,
\phi(t_{\overline{n}})]  \rangle \;.
\end{eqnarray}
\narrowtext
\section{Conclusions}

In this paper we have addressed the question: what kind of thermal
functions in RTF emerge by analytic continuations of the $n$-point
thermal Green functions defined in ITF. This amounts to perform
analytic continuations in the energies of the external legs from the
discrete imaginary values to real continuous ones.

The thermal Green functions are defined on a path in the complex
time plane, a path which is to a large extent arbitrary. On the
basis of this observation, we have carried out the above mentioned
continuations in the most straightforward and familiar manner by
deforming the contour, starting from the one that defines ITF to the
one defining RTF. In this way, we show that ITF $n$-point Green
functions become retarded or advanced thermal Green functions.

The results obtained in this paper, being exact and valid
independent of the approximation used in actual mainly perturbative
calculations, are partly discussed in \cite{eva2}, and we expect
that our approach gives additional information on the underlying
relations between ITF and RTF.


\newpage
\centering{FIGURE CAPTIONS} \\
FIG.1. The contour $C_I$ in the complex time plane, which defines
ITF. \\
FIG.2. The contour $C_R=C_3 \oplus C_1 \oplus C_2 \oplus C_4$ in
the complex time plane; the segments $C_1$ and $C_2$ lie on the
real axis, and $t_I = \Re t_0$.


\begin{references}
\bibitem[17]{foo}The modifications are as follows \cite{lan,nie}:
$G_{11}^{(0)}$ $\to$ $G_{++},\;$ $G_{12}^{(0)}$ $\to$ $G_{+-}, \;$
$G_{22}^{(0)}$ $\to$ $G_{--}$ and $G_{21}^{(0)}$ $\to$ $G_{-+}$.
Here $G$'s are given in (12) and (13) below.
\end{references}
\end{document}